\documentclass[aps,onecolumn,prd,showpacs,showkeys,preprintnumbers,superscriptaddress,nobibnotes,notitlepage,floatfix,longbibliography,nofootinbib]{revtex4-2}

\usepackage{graphicx}
\usepackage{amsmath}
\usepackage{amssymb}
\usepackage{xcolor}
\usepackage[colorlinks=true,citecolor=blue,urlcolor=blue]{hyperref}
\usepackage{enumitem}

\def\be{\begin{equation}}
\def\ee{\end{equation}}
\def\bea{\begin{eqnarray}}
\def\eea{\end{eqnarray}}

\def\mev{\, {\rm MeV}}

\def\ev{\, {\rm eV}}

\newcommand{\gsim}{\lower.7ex\hbox{$\;\stackrel{\textstyle>}{\sim}\;$}}
\newcommand{\lsim}{\lower.7ex\hbox{$\;\stackrel{\textstyle<}{\sim}\;$}}

\begin{document}


\title{Clustering with Light (but Massive) Relics}

\author{Jason Kumar}
\affiliation{Department of Physics and Astronomy, University of Hawai'i, Honolulu, HI 96822, USA}

\author{Pearl Sandick}
\affiliation{Department of Physics and Astronomy, University of Utah, Salt Lake City, UT  84112, USA}

\author{Shuting Xu}
\affiliation{Department of Physics and Astronomy, University of Hawai'i, Honolulu, HI 96822, USA}

\begin{abstract}
We consider the effect of Light (but Massive) Relics (LiMRs) on the clustering of matter in the early
Universe.  We account for the fact that LiMRs which are massive enough may cluster on large length
scales at early times, and may thus impact weak lensing of the cosmic microwave background (CMB) even
on small angular scales.  In particular, we find that LiMRs in the $\gtrsim \ev$ mass range (and even
$> 10~\ev$), can constitute a non-negligible component of dark matter.  This opens up a class of scenarios
in which energy is injected as dark radiation, but begins to redshift as
matter before recombination, thus avoiding constraints on $\Delta N_{eff}$ while providing an
$\ev$-range dark matter component.
\end{abstract}

\maketitle

\section{Introduction}

There are a wide variety of scenarios for beyond the Standard Model (BSM) physics in which
new relativistic particles are produced in the early Universe.  If such particles behave as
radiation up to the epoch of recombination, then they will contribute to the number of effective
neutrinos ($N_{eff}$), and can be constrained by CMB observables~\cite{Planck:2018vyg}.
But if these particles, though
light, are not too light, then their energy density may redshift like matter in current epoch.
Such light massive relics may thus constitute a fraction of the dark matter abundance, and may have
done so even at the epoch of recombination.  A key way to constrain such Light (but Massive) Relics (LiMRs)~\cite{Xu:2021rwg}
is to examine their effect on the growth of large scale structure.  In particular, LiMRs which are
sufficiently fast can escape from the gravitational potentials generated by matter perturbations,
effectively suppressing the clustering of matter into structures.  In this work, we will perform a
general study of the impact on matter clustering of LiMRs in the $\sim 0.01 - 20~\ev$ mass range.

It has long been known that light particles can suppress the formation of structures on scales smaller
than their free-streaming scale~\cite{Bond:1980ha} (see Refs.~\cite{Baumann:2022mni,Green:2022bre} for reviews, as well
as Ref.~\cite{Craig:2024tky}).
Observations of large scale structure thus provide a key constraint on scenarios of warm/hot dark matter~\cite{Planck:2018vyg,Lin:2023fao}.
Similarly, the mass of neutrinos can be constrained by the extent to which neutrinos would suppress
the growth of matter perturbations, which in turn leads to a suppression of the weak gravitational lensing
of the Cosmic Microwave Background (CMB) by matter perturbations.  However, studies of the effect of
LiMRs on clustering typically focus on a global analysis of cosmological constraints of specific models, often
making assumptions which limit the genericity of the analysis.  For example, an analysis in
Ref.~\cite{Xu:2021rwg} provided constraints on models of LiMRs, but under specific assumptions regarding
the effective temperature of the LiMRs, and under the assumption that they did not cluster on any scale.

By contrast, here we focus on the effect of LiMRs on matter clustering for a generic choice
of particle mass and energy density.  For this purpose, we will perform a linearized analytic analysis with some
well-motivated approximations.  In particular, we will account for the fact that a LiMR which
is heavy enough can cluster on large length scales.  If such clustering occurs at early times, it can impact
weak lensing of the CMB, even on small angular scales.
We will find that LiMRs with a mass $\gtrsim \ev$ can constitute a non-negligible fraction of the matter
density without significantly suppressing the weak lensing of the CMB.
This result broadens the allowed scenarios for the beyond-the-Standard Model (BSM) physics in the
early Universe.  A dark sector energy density which redshifts as radiation well before recombination, and would
thus be constrained by its contribution to $\Delta N_{eff}$, may instead redshift as matter at the time of
recombination, and could be an unconstrained component of dark matter.

The plan of this paper is as follows.  In Section~\ref{sec:LiMRs}, we will describe the impact of LiMRs
on the clustering of matter and the weak lensing of the CMB, in the linearized approximation.
In Section~\ref{sec:Results}, we will present our results describing the region of parameter space in
which LiMRs can consititute a non-negligible fraction of dark matter, without significantly suppressing
weak lensing of the CMB.  We will conclude in Section~\ref{sec:Conclusion}.

\section{LiMRs and the growth of matter perturbations}
\label{sec:LiMRs}

For simplicity, we will assume that the dark sector consists of a single self-conjugate spin-0 particle $X$ with mass $m_X$.
We will also assume that, at some time between the epoch when positrons annihilate away and recombination, $X$
has thermalized with an effective temperature $T_X = r T_\gamma$, where $T_\gamma$ is the temperature of the
photon bath, and $r$ is a parameter which specifies the cosmological initial conditions.  Since we assume that
$X$ and the photon-baryon plasma are both subsequently decoupled, with effective temperatures which decrease with
the expansion of the universe, the ratio of temperatures remains constant up to the current epoch.

We assume that $m_X$ is large enough that $X$ redshifts like matter today.  Since the number density of $X$ scales as
$a^{-3}$, the ratio of the energy density of the
dark sector to the matter density can then be written as
\bea
f_X = \frac{\rho_X}{\rho_M}
&=& 10^{-2} \frac{m_X}{0.075 ~\ev} r^3 g .
\label{eq:fX}
\eea
where $g=1$ in the case of a real scalar particle, which we consider here.
We would have $g=2, 3/2, 3$ for the cases of a complex scalar, Majorana fermion, or Dirac fermion, respectively
(henceforth, we simply set $g=1$).
For all models which we consider, we will find $f_X \ll 1$, justifying an expansion only to
linear order in $f_X$.

\subsection{The Growth of Matter Perturbations}

We follow the formalism used in Ref.~\cite{Craig:2024tky} (see also Refs.\cite{Baumann:2022mni,Green:2022bre}).
In the absence of perturbations, we can describe the universe with a Friedmann-Robertson-Walker metric
\bea
ds^2 &=& dt^2 - a(t)^2 [ dx^2 + dy^2 + dz^2] = a(\eta)^2 [d\eta^2 - (dx^2 + dy^2 + dz^2)] ,
\eea
where $\eta$ is conformal time ($d\eta / dt = 1/a$).
We will set $a_0 = a(t_0) = 1$, where $t_0$ is the current age of the Universe.
We will focus on the epoch after recombination,
in which the stress-energy tensor is matter-dominated, and $a \propto t^{2/3}$,
$\eta (t) = 3 t_0 \left( 1 + z \right)^{-1/2} \propto t^{1/3}$.\footnote{Dark energy
will effect cosmological evolution at late redshifts ($z \sim 1$), but
this will not significantly effect our results, and for simplicity we ignore it here.}

The density contrast of the $i$th particle species is defined as $\delta \rho_i / \bar \rho_i$,
where $\bar \rho_i$ is the unperturbed energy density and $\delta \rho$ is the perturbation.
Defining $\delta_i (k)$ as the Fourier transform of the density contrast, the equation of motion for $\delta_i$ can be
found by expanding Einstein's equation to leading order in the perturbations, yielding
\bea
\frac{d^2\delta_i}{dt^2} + 2H \frac{d\delta_i}{dt} &=& -\frac{k^2}{a^2} c_i^2 \delta_i +
\frac{3}{2}H^2  \sum_j f_j \delta_j ,
\label{eq:EOM}
\eea
where  $H = \dot a / a$ is the Hubble parameter, $\sum_i f_i =1$, and
$c_i$ is the propagation speed of species $i$.  Since $X$ decouples while
relativistic, we have
\bea
c_X &=& \frac{\langle p_X \rangle}{m_X} =\frac{3 T_X}{m_X} = \frac{3r}{m_X} \frac{T_{0\gamma}}{a(t)} ,
\nonumber\\
&=& 0.014 r \left(\frac{m_X}{0.05~\ev} \right)^{-1} (1+z) \propto t^{-2/3} ,
\label{eq:cX}
\eea
where $T_{0\gamma}$ is the current temperature of the photon bath.  A rough measure of when particle species
$X$ redshifts like matter is when $c_X \lesssim 1$.

We see from the right-hand side of eq.~\ref{eq:EOM} that growth of density perturbations of the $i$th
species are suppressed by the pressure of that species, but are driven by the gravitational potential
generated by perturbations of all matter species.
To characterize the effect of non-trivial pressure, we introduce the parameter
$\alpha_i\equiv (3/2) (k^2 c_i^2 t^2 / a^2) = k^2 / k^2_{i,{\rm fs}}$,
where $k_{i,{\rm fs}} = \sqrt{2/3} (a/c_i t)$ is the redshift-dependent free-streaming wavenumber for particle $i$.  Using the
fact that $H \approx 2/3t$ in the matter-dominated era, we can rewrite the equation of motion as
\bea
\frac{d^2\delta_i}{dt^2} + \frac{4}{3t} \frac{d \delta_i}{dt} &=& -\frac{2}{3t^2} \alpha_i  \delta_i +
\frac{2}{3 t^2}  \sum_j f_j \delta_j .
\eea

If $k \gg k_{i,{\rm fs}}$, then $\alpha \gg 1$.  We can simplify the solution to the equation of motion
by approximating $\alpha$ as a constant,  The solution for $\delta_i$ is a damped oscillatory solution, with
a damping envelope $\propto t^{-1/6}$.  This is the expected result, namely, that on length scales much smaller
than the free-streaming scale, density fluctuations are erased as particle free-stream out of the gravitational
potentials.  Since we will not need the details of the damping solution, we are justified in ignoring the time-dependence
of $\alpha$; it is sufficient to note that the density contrast will rapidly shrink to zero for wavenumbers larger
than the free-streaming scale.

On the other hand, if $k \ll k_{i,{\rm fs}}$, then $\alpha \ll 1$.  Let us define
$f_{cl} = \sum_i f_i$, $f_{cl }\delta_{cl} = \sum_i f_i \delta_i$,  where the sums are over particles
for which $\alpha_i \ll 1$.  In other words, $f_{cl}$ is the fraction of matter which clusters, and
does not free-stream at wavenumber scale $k$.  The fraction of matter which does free-stream at scale
$k$ is defined as $f_{fs} = 1-f_{cl}$.
Combining the equations of motion for all clustering matter (and taking $\delta_j \rightarrow 0$ for all
$j$ for which $k \gg  k_{j,{\rm fs}}$), we then find
\bea
\frac{d^2\delta_{cl}}{dt^2} + \frac{4}{3t} \frac{d\delta_{cl}}{dt} &=&
\frac{2}{3t^2} (1- f_{fs}) \delta_{cl} ,
\label{eq:cl}
\eea
which can be solved with the ansatz $\delta_{cl} \propto t^\gamma$.  There are two solutions,
and the solution which grows with time is given by $\gamma \sim (2/3) -(2/5) f_{fs} +{\cal O}(f_{fs}^2)$.  If there
were no free-streaming matter, then matter perturbations would grow as $t^{2/3}$.

We thus see that the presence of matter which free-streams at scale $k$ has two effects.  First, it reduces
the fraction of matter which clusters (that is, whose density contrast grows with time) by a factor
$1-f_{fs}$.  Secondly, it reduces
the rate by which the density contrast of clustering matter grows.

Since the free-streaming scale is redshift-dependent, whether or not $X$ free-streams will depend both
on $t$ and on the scale $k$.
Suppose the particle $X$ began to redshift like matter at time $t_X$, and that between times $t_X$ and
$t> t_X$, we have $k \gg  k_{X,{\rm fs}}$.  That is, between times $t_X$ and $t$, $X$ particles free-stream
on wavenumber scale $k$.  We then find that the growth of $\delta_{cl}(k) $ is suppressed by a factor
$(t/t_X)^{-(2/5) f_X} = 1 - (2/5) f_X \ln (t/t_X) + {\cal O}(f_X^2)$.

\subsection{The Weak Convergence Lensing Power Spectrum}

We can best measure the matter power spectrum through the weak gravitational lensing of the CMB by matter perturbations
(see Ref.~\cite{Lewis:2006fu} for a review).
The quantity of interest for us then  is actually the weak convergence lensing power spectrum, which is related to
the power spectrum of the integrated gravitational potential generated by the matter perturbations.
If we can ignore any suppression of the clustering of matter perturbations due to free-streaming, then the unsuppressed
CMB lensing convergence power spectrum is  given in the Limber approximation by~\cite{Lewis:2006fu,Craig:2024tky}
\bea
C_{\ell}^{\kappa \kappa} &\simeq& 2\pi^2 \ell \int_{\eta_*}^{\eta_0} d\eta~\eta
\left[\frac{9\Omega_m^2(\eta) {\cal H}^4(\eta)}{8\pi^2} \frac{P(\ell/(\eta_0-\eta);\eta)}{\ell/(\eta_0-\eta)} \right]
\nonumber\\
&\,& \times \left(\frac{\eta_* -\eta}{(\eta_0 -\eta_*)(\eta_0 -\eta)} \right)^2 ,
\eea
where ${\cal H} \equiv (1/a) da/d\eta = aH$ is the conformal Hubble parameter,
$P(k;\eta)$ is the matter power spectrum at conformal time $\eta$,
and $\eta_{*(0)}$ is the conformal
time at recombination (the current epoch).
The Limber approximation is accurate at large $\ell$.  In this approximation, the comoving wavenumber scale
$k$ which contributes to weak lensing at angular scale $\ell$ at conformal time $\eta$ is given by
$k \sim \ell / (\eta_0 - \eta)$.\footnote{In this approximation,
lensing at any conformal time is dominated by perturbations whose wavenumber is transverse to the line of sight.}

We initially consider the case in which there is no free-streaming matter, and no suppression to the clustering
of matter perturbations.  Since linear gravitational potential perturbations do not grow in the matter dominated
epoch~\cite{Baumann:2022mni,Green:2022bre}, we do not expect the convergnece
lensing power spectrum to grow either.  We can see this by noting that the
primordial matter perturbations (absent any suppression from free streaming) grow like $t^{2/3}$, leading $P(k;\eta)$
to grow as $t^{4/3}$, while ${\cal H}^4 \propto t^{-4/3}$.
We thus find that ${\cal H}^4 P(k;\eta) \propto P^0 (k)$, where $P^0 (k)$ is the primordial matter power
spectrum.\footnote{Note, there is a suppression in the growth of perturbations for modes which only re-entered the
horizon well after recombination, and thus have not had much time to grow.  But these modes do not significantly
effect the weak lensing power spectrum at high-$\ell$, so we will ignore this suppression.}

Choosing $P^0 (k) \propto k^{-3}$, we find
\bea
C_{\ell}^{\kappa \kappa}
&\propto& \int_{\tilde \eta_*}^{\tilde \eta_0} d\tilde \eta~\tilde \eta ~
(\tilde \eta_* -\tilde \eta)^2 (\tilde \eta_0 -\tilde \eta)^2  ,
\eea
where we define a dimensionless conformal time $\tilde \eta = \eta / (3t_0)$.
$\tilde \eta_0 =1$ is the dimensionless conformal time now, and
$\tilde \eta_* = (1+z_*)^{-1/2}
\sim 0.03$ is the dimensionless
conformal time at recombination.

\subsection{Linearized approximation to the suppression of density perturbations}

 We will adopt the simplifying assumption that we can set $\alpha_X =0$ whenever
$\alpha_X < 1$ (that is, $X$ clusters just
like cold dark matter), and can make the approximation $\alpha_X \gg 1$ whenever $\alpha_X > 1$.  In
other words, we will assume that there is a sharp transition between the scales on which $X$ clusters like cold dark matter (CDM) and
those on which it
free-streams.  To linear order in $f_X$, we can then write
\bea
C_{\ell}^{\kappa \kappa}
&\propto& \int_{\tilde \eta_*}^{\tilde \eta_0} d\tilde \eta~\tilde \eta ~
(\tilde \eta_* -\tilde \eta)^2 (\tilde \eta_0 - \tilde \eta)^2 \times F(\tilde \eta ; \ell)  ,
\eea
where
\bea
F(\tilde \eta ; \ell ) &=& \exp\left[\ln \left(1 - 2 f_X
- \frac{12}{5} f_X \ln \frac{\tilde \eta}{\max[\tilde \eta_X , \tilde \eta_*]} \right)
\times \theta(\tilde \eta - \tilde \eta_X) \times \theta \left(\frac{\ell}{\tilde \eta_0 - \tilde \eta} - \tilde k_{fs} (\tilde \eta) \right)  \right] ,
\nonumber\\
\label{eq:F}
\eea
and
\bea
\tilde \eta_X &=& \left[  71.4 \left(\frac{m_X /r}{0.05~\ev} \right) \right]^{-1/2} ,
\nonumber\\
\tilde k_{fs } (\tilde \eta) &=& 174~\frac{ m_X /r}{0.05~\ev}  \tilde \eta .
\eea
$\tilde \eta_X$ is the dimensionless conformal time at which $X$ begins to redshift as matter,
and $\tilde k_{fs} = (3t_0) k_{fs}$ is the dimensionless co-moving wavenumber above which $X$ free-streams.
The first Heaviside step-function  in eq.~\ref{eq:F} forces
$F(\tilde \eta) \rightarrow 1$ when $X$ is relativistic, since it then does not contribute to the
density of free-streaming matter.\footnote{If $X$ is relativistic at early times, then it will contribute to
$\Delta N_{eff}$ and to a suppression of
$\Omega_m$.  We will address the observational impact of these effects in more detail later on.}
Similarly, the second Heaviside step-function  forces
$F(\tilde \eta) \rightarrow 1$ when $k < k_{fs}$, since on those scales $X$ clusters, and does not
contribute to the suppression of the matter power spectrum.  Note that even though
$k_{fs} (\tilde \eta) \propto \tilde \eta$, we also have $\tilde k \propto (\tilde \eta_0 - \tilde \eta)^{-1}$,
implying that, for any fixed choice of angular scale $\ell$, at sufficiently late times
(as $\tilde \eta \rightarrow \tilde \eta_0$) $X$ will free-stream and suppress the lensing power spectrum.
This essentially amounts to the fact that, in the Limber approximation, lensing is dominated by wavenumbers
transverse to the line of sight, and at increasingly late times a fixed angular scale will be spanned by
increasingly small transverse comoving length scales.

The term with $-2 f_X$ arises because,  on scales at which $X$ free-streams, the fraction of matter
which clusters is suppressed by $1-f_X$, yielding a suppression of the potential power spectrum
$\propto 1 - 2 f_X$.  Similarly, if $X$ free-streams on wavenumber scale $k$, then the growth of
matter perturbations is suppressed by a factor $1- (6/5) f_X \ln (\tilde \eta / \tilde \eta_X)$
(assuming $\tilde \eta_X > \tilde \eta_*$),
yielding a suppression of the potential power spectrum
$\propto 1- (12/5) f_X \ln (\tilde \eta / \tilde \eta_X)$.  But even if $X$ is heavy enough to redshift like
matter during the radiation-dominated epoch, it will only suppress matter perturbations after the Universe
becomes matter-dominated.  Up to a small correction, we can approximate this by replacing $\tilde \eta_X$
with $\max [\tilde \eta_X, \tilde \eta_*]$, where $\tilde \eta_*$ is the dimensionless conformal time of recombination.

We are interested in the fractional shift in the convergence lensing power spectrum due to the
free-streaming of $X$, which may be expressed by the ratio
\bea
R (\ell) &=& \frac{C_{\ell}^{\kappa \kappa} - \left(C_{\ell}^{\kappa \kappa}\right)_{F=1}}
{\left(C_{\ell}^{\kappa \kappa}\right)_{F=1}} .
\eea
Note that we have not included the matter power spectrum suppression due to other particle species
which do not cluster, such as neutrinos.  This treatment is justified because we are working to linear
order in the mass fraction of all non-clustering particles (that is, $f_i \ll 1$ for all particle species
$i$ which do not cluster).  As such, the contributions to $R(\ell)$ from neutrinos would simply add
to the contribution from $X$, up to ${\cal O}(f^2)$.

\begin{figure*}[t]
    \centering
    \includegraphics[width=0.8\textwidth]{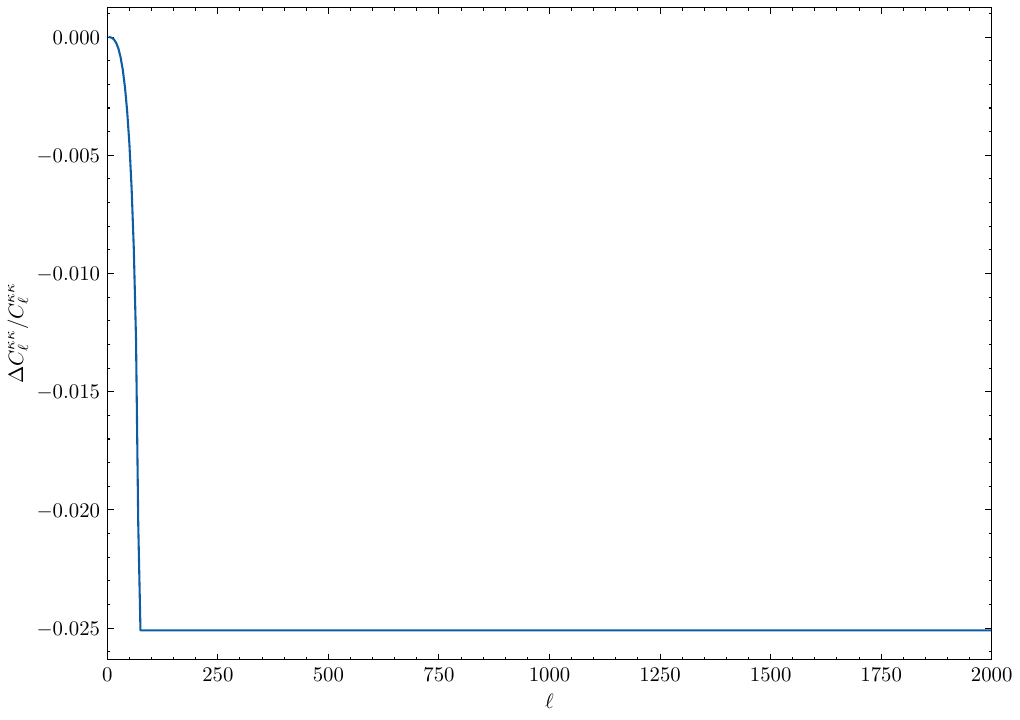}\\
    \caption{Plot of $\Delta C_{\ell}^{\kappa \kappa} / C_{\ell}^{\kappa \kappa}$ as a function
    of $\ell$, setting $f_\nu = 4 \times 10^{-3}$ and $\sum_\nu m_\nu = 0.058~\ev$ ($g = 3/2, n_s=1$).}
    \label{fig:Lensing}
\end{figure*}

To understand the behavior of $R(\ell)$, it is useful to apply this formalism to the case of a massive
neutrino.  For this benchmark case, we will set $r=(4/11)^{1/3}$ (taking $g=3/2$) and $m_\nu = 58~\mathrm{meV}$, which is the
lower bound on the mass of the heaviest neutrino, consistent with oscillation experiments~\cite{ParticleDataGroup:2024cfk}.
We plot $R(\ell)$ in
Figure~\ref{fig:Lensing}.  We see that $R(\ell) \rightarrow \sim -0.025$ at large $\ell$.  Indeed,
$R(\ell)$ is constant for $\ell \gtrsim 100$.  The reason is because, for this benchmark, the neutrino
redshifts as matter for $\tilde \eta > \tilde \eta_\nu \sim 0.1$.  For this range of $\tilde \eta$, one always
finds $\ell / (\tilde \eta_0 - \tilde \eta) > \tilde k_{fs} (\tilde \eta)$ for $\ell > 100$.
In other words, the massive neutrino free-streams for all conformal times relevant for lensing on
an angular scale $\ell > 100$.  Since $\ell$ only enters the expression for $R(\ell)$ in the Heaviside
step function which forces the suppression to zero when the neutrino does not free-stream, this implies
that $R(\ell)$ should be independent of $\ell$ once $\ell > 100$.

Note that, although the asymptotic value of $R(\ell)$ is consistent with the asymptotic value found in
Figure 2 of Ref.~\cite{Craig:2024tky}, the detailed shape at smaller $\ell$ is discrepant.  This is
related to our choice of denominator in defining $R(\ell)$.  In particular, we have adopted a very simple
cosmology in which we have set $\Omega_m = 1$ for all times after recombination.  The convergence lensing power
spectrum for the case of a massive neutrino is then compared to that of a model with $\Omega_m =1$, but where
no particles free-stream.  The numerical analysis of  Ref.~\cite{Craig:2024tky} adopts a more realistic
cosmology in which $\Omega_m \sim 0.3$ at very late times, though $\Omega_m \sim 1$ for most of the times
relevant for CMB lensing.  For large $\ell$, where a large range of conformal times contribute to weak lensing,
the deviation of our analytic treatment from a full numerical calculation is negligible.  The discrepancy becomes
larger at smaller $\ell$, for which lensing suppression only arises at late times.  In that case, the precise
result also depends in detail on the model choices which one makes (for example, whether setting $m_\nu =0$ results
in a reduction of $\Omega_m$, or in a reduction in $\Omega_{cdm}$ while keeping $\Omega_m$ fixed).  Since our
interest is not in a global analysis of a specific model, but a general analysis of the effect of LiMRs on weak
lensing, our analytic treatment will be sufficient.

We can now consider from an analytic point of view how $R(\ell)$ changes as we change $m_X$ and $r$ (or,
equivalently, $f_X$).
If we change $m_X$ while holding $m_X / r$ fixed, then the range of conformal times over
which the growth of matter perturbations is suppressed remains unchanged.  But we then have
$f_X \propto m_X r^3 \propto m_X^4$, implying that larger masses will lead to a larger suppression
of the lensing power spectrum.

On the other hand, if we increase $m_X$ while keeping $f_X$ fixed (that is, taking $r \propto m_X^{-1/3}$),
then magnitude of $F$ remains large fixed (up to logarithmic terms), but $\tilde k_{fs} (\tilde \eta) \propto m_X^{4/3} \tilde \eta$,
while $\tilde \eta_X \propto m_X^{-2/3}$.  So even though $X$ begins to redshift as matter at earlier times, its
free-streaming wavenumber is larger, implying that for fixed angular scale $\ell$, the free-streaming of $X$ only
suppresses the convergence lensing power spectrum at later times.  To see the effect of the suppression of the
integrated power spectrum over all conformal times after $X$ becomes non-relativistic, one must go to larger $\ell$.

\begin{figure*}[t]
    \centering
    \includegraphics[width=0.8\textwidth]{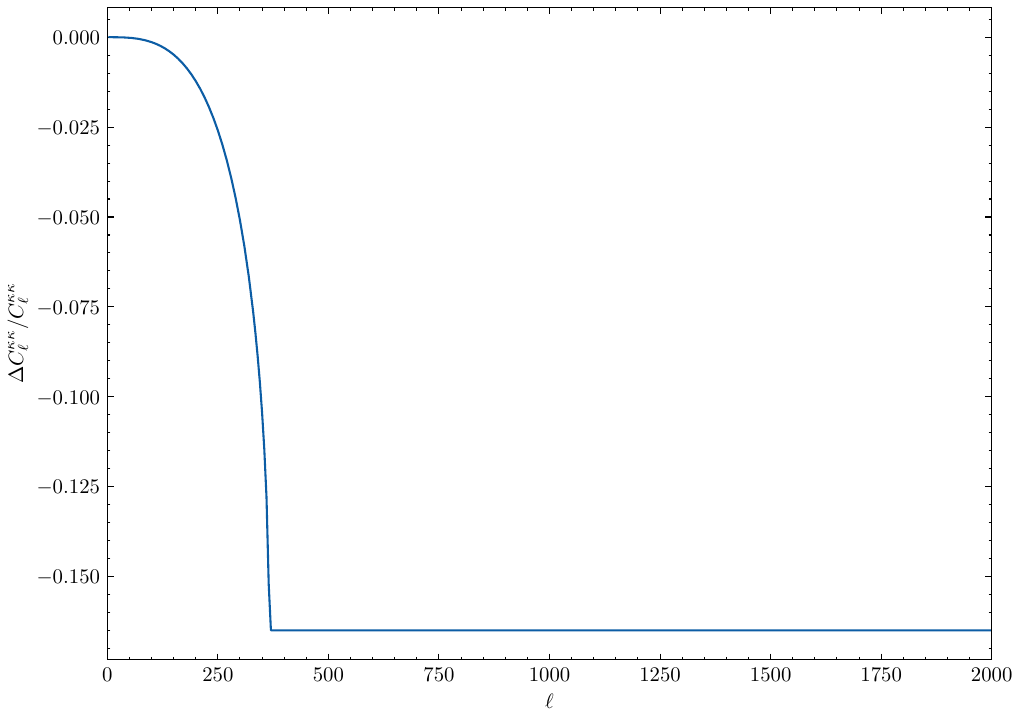}\\
    \caption{Plot of $\Delta C_{\ell}^{\kappa \kappa} / C_{\ell}^{\kappa \kappa}$ as a function
    of $\ell$, setting $r = (4/11)^{1/3}$ and $\sum_\nu m_\nu = 0.3~\ev$ ($g = 3/2, n_s=1$).}
    \label{fig:HeavyNu}
\end{figure*}

As a second benchmark, we can consider the case of a heavy neutrino with $m_\nu = 0.3~\ev$, which is roughly the maximum
neutrino mass allowed by primary observations of the CMB~\cite{Planck:2018vyg}.
In this case, we find $f_\nu \sim 0.02$, with $R(\ell)$ shown in
Figure~\ref{fig:HeavyNu}.  At large $\ell$, we find $R(\ell) \rightarrow -0.17$.  As the contribution of massive neutrinos
to $R(\ell)$ can span the range from $-0.025$ to $-0.17$ while remaining largely consistent with cosmological observations,
we will treat those values as a benchmarks for a scenario in which $X$ is not ruled out by CMB weak lensing measurements.

We emphasize, of course, that this does not represent a global analysis of any particular model against all cosmological observables.
It is merely a simple way of quantifying the effect of a LiMR on weak lensing measurements.  An interesting application of this
analysis is to the case
in which $\eta_X < \eta_*$, and $X$ redshifts as matter at recombination and does not contribute the $\Delta N_{eff}$.  In this
case, the dominant constraint on $X$ from cosmology will be on the growth of structures, and the effect of free-streaming can be well
quantified by $R(\ell)$.

\section{Results}
\label{sec:Results}

\begin{figure*}[t]
    \centering
    \includegraphics[width=0.8\textwidth]{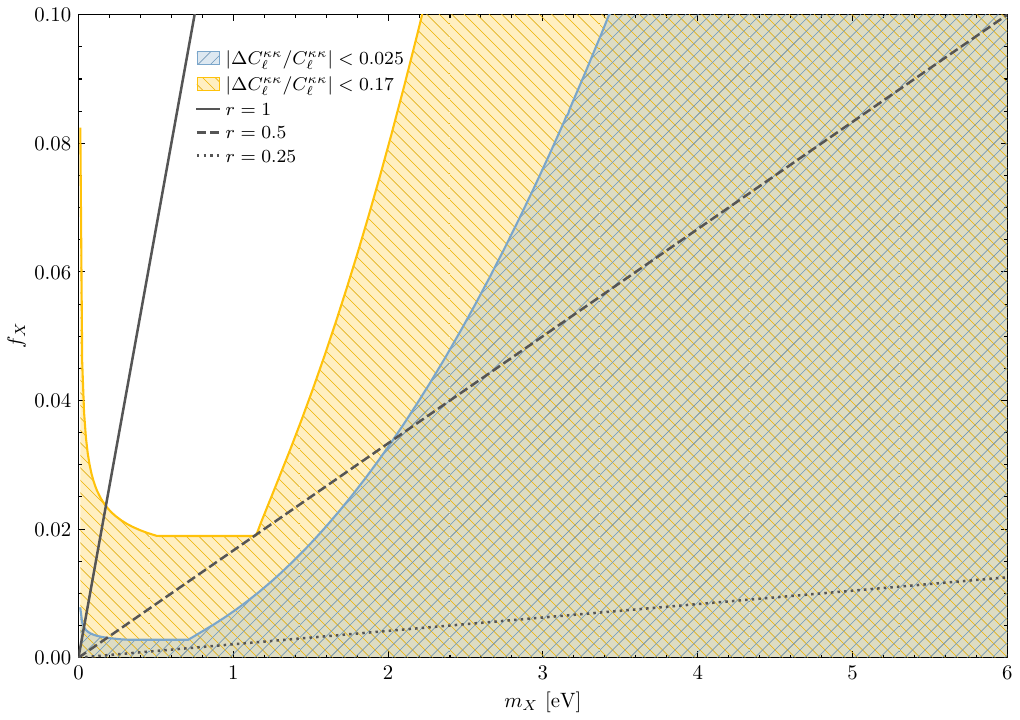}\\
    \caption{Regions of the $(m_X, f_X)$ parameter space in which $|\Delta C_{\ell}^{\kappa \kappa} / C_{\ell}^{\kappa \kappa}|
    < 0.025~(0.017)$
    for $\ell \leq 2000$ are plotted in blue (yellow).  Also plotted are contours of constant $r \equiv T_X / T_\gamma$, as labelled.}
    \label{fig:MainPlot}
\end{figure*}

Using the procedure described in the previous section, we plot in Figure~\ref{fig:MainPlot} the region in the $(m_X , f_X)$-plane for
which $|\Delta C_{\ell}^{\kappa \kappa} / C_{\ell}^{\kappa \kappa}| < 0.025$  for $\ell \leq 2000$ (blue region).
In this region, the effect of clustering on $X$ is less than that of the lightest massive neutrino consistent with
oscillation experiments.  We also plot the region in the $(m_X , f_X)$-plane for
which $|\Delta C_{\ell}^{\kappa \kappa} / C_{\ell}^{\kappa \kappa}| < 0.17$ (yellow region).  In this region of parameter space,
the effect of $X$ on clustering is less than that of the most massive neutrino which is consistent with primary
CMB observations.

The shape of these contours can be understood qualitatively from the considerations of the previous section, by considering
the limit in which we increase $m_X$, while keeping $f_X \propto m_X r^3$ fixed.  When $m_X$ is sufficiently small,
$X$ already free-streams on small angular scales (that is, large $\ell$) as soon as it becomes non-relativistic.  In that limit, when
$m_X$ increases, $X$ becomes non-relativistic (and begins to suppress structure) at earlier times, leading to a larger
value of $|\Delta C_{\ell}^{\kappa \kappa} / C_{\ell}^{\kappa \kappa}|$.
This leads to the negative slope for the contours at small $m_X$.
But once $m_X$ is large enough, $X$ is already
non-relativistic at the time of recombination; at this point, further increasing $m_X$ does not increase the suppression of weak lensing.
This leads to the flat feature seen in the contours in Figure~\ref{fig:MainPlot}.

But at a fixed angular scale $\ell$ and
for sufficiently large values of $m_X$, $X$ begins to cluster like cold dark matter after it becomes non-relativistic,
and only free-streams at sufficiently late conformal time (when the relevant length scale is sufficiently
small).  In this limit,
we find that even though increasing $m_X$ increases the amount of time over which $X$ redshifts as matter, it decreases the amount
of time over which $X$ free-streams, resulting in a smaller value of $|\Delta C_{\ell}^{\kappa \kappa} / C_{\ell}^{\kappa \kappa}|$.
This effect leads to the positive slope for the contours at sufficiently large $m_X$,
which appears only because we have accounted for the clustering of LiMRs at early conformal times.

At very small mass, the values of $f_X$ consistent with a small suppression of weak lensing become relatively large,
which seems counterintuitive.  But it it should be noted that at small $m_X$, for such large values of $f_X$, one
finds $r>1$ (that is, the effective temperature of $X$ would be larger than that of the photons).  Although
most models which are typically studied have $r<1$, cosmological histories with $r>1$ are also possible.  In any case,
Figure~\ref{fig:MainPlot} only describes the effect of LiMRs on weak lensing on the CMB.  For such small values of
$m_X$, when $X$ is relativistic at the time of recombination, it will also contribute to $N_{eff}$.  As a result,
there will be more detailed constraints from other CMB observables, but which are very specific to the detailed model, and
beyond the scope of this work.

Note that, in Figure~\ref{fig:MainPlot}, we have only plotted
regions of parameter space for which $f_X \leq 0.1$, since we are using a linearized approximation.  But we expect larger values of $f_X$ to be allowed for larger $m_X$.\footnote{Although we Figure~\ref{fig:MainPlot} extends only to 6 eV for plotting purposes, the result extends to
larger mass.}  This simply reflects the fact that as $m_X$ becomes sufficiently large,
$X$ can be treated essentially
as cold dark matter, and clusters as such.  In that sense, our analysis smoothly connects (though only at linear order) the cosmological effects
of hot dark matter, which does not cluster, and cold dark matter, which does.

Finally, we consider the impact of using only a linearized analytic approximation, rather than a global numerical
analysis of all cosmological observables.  For any particular model, a global analysis will provide significantly
tighter constraints than a linearized analysis of CMB lensing alone.  Indeed, a global analysis of data including
DESI measurements~\cite{DESI:2024mwx} of baryon acoustic oscillations (BAO) indicates a preference for a summed neutrino mass which is
smaller than the minimum value consistent with neutrino oscillation experiments.  This indicates an excess of clustering
beyond what one would expect in standard cosmology, which may be interpreted in a specific model as an absence of the
suppression arising from neutrino free-streaming.
As discussed in Ref.~\cite{Craig:2024tky}, for example, the impact of DESI data on this results arises not so much from
the measurement of the matter power spectrum (which is more precisely measured from CMB weak lensing), but from
DESI's precise measurement of $\Omega_m h^2$.  Since $C^{\kappa \kappa}_{\ell} \propto (\Omega_m h^2)$, a small shift
in $\Omega_m h^2$ would be sufficient to mimic (or cancel) the suppression of clustering arising from massive
neutrinos with $\sum_\nu m_\nu = 58~\mathrm{meV}$.  Recall that, in our simplified analysis, we assumed matter
domination, which amounts to setting $\Omega_m = 1$ after recombination.
Thus, it is clear that we are ignoring effects and datasets which can effect the consistency
of any particular model with observational data.

However, this example also illustrates the utility of our approach.  Our goal is not to test any particular model
of early Universe cosmology, but rather to quantify the effect the LiMRs on the suppression of matter clustering.  For
this purpose, the global analyses of data (including DESI BAO observations) confirm what is suggested by the analytic
result; namely, the suppression of the CMB weak lensing power spectrum is well approximated by the
sum several effects, one of which is the free-streaming of massive neutrinos, though there are other effects which
are qualitatively similar.  Indeed, the fact that a global analysis of cosmological data shows some tension with
laboratory constraints on neutrino masses may suggest that there are indeed other contributions to the suppression
or enhancement of CMB weak lensing.  Our aim is not to test any particular model which may be consistent with all
of the data (though this is an active area of research), but rather to compare the contributions to CMB weak lensing
suppression of LiMRs and neutrinos in standard scenarios.   We have identified regions of the LiMR parameter space in
which the impact on CMB weak lensing suppression is comparable to that of standard neutrinos scenarios.  Whatever other
new physics may also impact CMB weak lensing, LiMRs in these regions of parameter space have no more of an effect
than what can be ascribed to our uncertainty in neutrino particle physics.

\section{Conclusions}
\label{sec:Conclusion}

We have considered the effects of Light (but Massive) Relics (LiMRs) on the weak lensing of the CMB, accounting for the growth
of their perturbations as a function of time.  In particular, we have accounted for the fact that LiMRs may initially cluster as
matter on comoving length scales which affect lensing at a particular angular scale, only to free-stream at a much later time,
when the relevant comoving length scales are smaller.

We have found that LiMRs with mass $\gtrsim \mathrm{few}~\ev$  can constitute a mass fraction of $f_X \sim 0.1$, while
suppressing clustering only to an extent similar to that of neutrinos in standard scenarios.  Indeed, we expect that somewhat
heavier LiMRs can constitute an even larger mass fraction, but we have limited our analysis to $f_X \leq 0.1$, since we have
performed an analytic analysis to linear order in $f_X$.  These results follow the natural intuition that, as the mass of a
LiMR becomes larger, its allowed mass fraction will also become larger, as it simply becomes cold dark matter.

In particular, these results suggest that a sizeable fraction of dark matter can be composed of $\sim \ev$-scale particles which
would have redshifted as radiation in the epoch well before recombination.  There are a variety of scenarios in which
relativistic dark sector particles are injected in the early Universe.  For example, relativistic dark sector particles may
be produced from a dark sector first-order phase transition (see, for example,~\cite{Dent:2025lwe}).  If this phase transition occurs
at a temperature in the $0.1 - 1~\mev$ range, then the associated gravitational wave signal may be detectable at nanohertz range
gravitational wave observatories.  But for this signal to be observable, one typically requires a large latent heat, implying
the injection of a large dark energy density.  These relativistic dark sector particles are
usually treated as dark radiation, and are tightly constrained by observational bounds on the number of effective neutrinos
($N_{eff}$) at recombination.
This tension complicates efforts to build models which can produce an observable nanohertz gravitional wave signal from
an FOPT, while remaining consistent with cosmological constraints.
However, if these dark sector particles have a mass in the $\gtrsim \ev$-range, then they may actually redshift
as matter by the time of recombination, alleviating these constraints on $\Delta N_{eff}$ while also yielding a ${\cal O}(\ev)$ dark
matter candidate.

It would be interesting to consider detection prospects for such a low-mass dark matter component. Recently,
Ref.~\cite{Dror:2024ibf} considered the indirect detection prospects for such a candidate.  Direct detection sensitivity to
bosonic dark matter in this mass range through absorption may potentially be obtained from the use of Dirac material~\cite{Hochberg:2017wce},
semiconductor~\cite{Bloch:2016sjj} and superconductor~\cite{Hochberg:2016ajh} targets.  In light of the motivation for this mass
range from cosmology, it may be worthwhile to revisit these and other detection strategies.

We have only considered the effects of LiMRs on clustering, and the associated impact on weak lensing of the CMB, using an
analytic treatment at linear order in $f_X$.  This is sufficient for the purpose of understanding the general effect
of LiMRs on matter clustering.  But it would be interesting to consider specific models and perform a global analysis,
accounting for deviations from matter domination, and corrections which are higher order in $f_X$.

{\bf Acknowledgements.}  We are grateful to Daniel Green and Barmak Shams Es Haghi for useful discussions.
JK and PS wish to acknowledge the Center for Theoretical Underground Physics and Related Areas (CETUP*), the Institute for Underground Science
at Sanford Underground Research Facility (SURF), and the South Dakota Science and Technology Authority for hospitality and financial support,
as well as for providing a stimulating environment.  JK would also like to acknowledge the University of Utah for its hospitality during the
completion of this work.
JK is supported in part by DOE grant DE-SC0010504.  The work of PS is
supported in part by the National Science Foundation
grant PHY-2412834.

\end{document}